\documentclass{aa}   
\usepackage{natbib}
\bibpunct{(}{)}{;}{a}{}{,} 
\usepackage{graphicx}
\usepackage{xcolor}
\usepackage{colortbl}
\usepackage{tabularx}
\usepackage{multirow}  
\usepackage{booktabs}  
\usepackage{caption}   

\usepackage[colorlinks,allcolors=blue]{hyperref}

\usepackage{comment}

\usepackage{hyperref}
\hypersetup{
    colorlinks=true,
    linkcolor=blue,
    citecolor=blue,
    urlcolor=blue,
    pdfborder={0 0 0}
}
\usepackage{cleveref}
\usepackage{txfonts}
\usepackage{tikz}
\usetikzlibrary{arrows.meta,shapes.arrows}
\usepackage{bbding}

\usepackage{orcidlink}
\usepackage{ulem}

\newcommand{\cg}[1]{{\bf \color{brown}#1}}

\begin{document}

   \title{Survey of (sub)mm water masers in low-mass star-forming regions}

   \author{P. K. Humire\orcidlink{0000-0003-3537-4849} \inst{\ref{inst.USP}}
   \and 
   C. Goddi\orcidlink{0000-0002-2542-7743}\inst{\ref{inst.USP},\ref{inst2},\ref{inst3}}
   \and G. N.\ Ortiz-León\orcidlink{0000-0002-2863-676X} \inst{\ref{inst.INAUE}}
  \and A. Hernández-Gómez\orcidlink{0000-0001-7520-4305} \inst{\ref{inst.TECH}}
  \and J-C Loison\orcidlink{0000-0001-8063-8685} \inst{\ref{inst.CNRS}}
    }

   \institute{\label{inst.USP}Departamento de Astronomia, Instituto de Astronomia, Geofísica e Ciências Atmosféricas da USP, Cidade Universitária, 05508-090 São Paulo, SP, Brazil,
   \label{email}pedrokhumire@usp.br
   \and Dipartimento di Fisica, Università degli Studi di Cagliari, SP Monserrato-Sestu km 0.7, I-09042 Monserrato (CA), Italy\label{inst2} 
   \and
INAF - Osservatorio Astronomico di Cagliari, via della Scienza 5, I-09047 Selargius (CA), Italy\label{inst3}
   \and \label{inst.INAUE}Instituto Nacional de Astrofísica, Óptica y Electrónica, Apartado Postal 51 y 216, 72000 Puebla, México
   \and \label{inst.TECH}Tecnologico de Monterrey, Escuela de Ingeniería y Ciencias, Avenida Eugenio Garza Sada 2501, Monterrey 64849, Mexico
   \and \label{inst.CNRS} ISM, Université de Bordeaux - CNRS, UMR 5255, F-33400 Talence, France
   }


 
  \abstract
    {Water masers are common in star-forming regions (SFRs), with the 22.235~GHz transition widely detected in both high- and low-mass protostars. In contrast, (sub)millimeter water maser transitions remain poorly studied, especially in low-mass SFRs, due to atmospheric limitations and lack of systematic surveys.}
      {We search for millimeter water masers in a sample of low-mass SFRs previously known to exhibit 22~GHz emission. Specifically, we target the $3_{1,3} - 2_{2,0}$, $10_{2,9} - 9_{3,6}$, and $5_{1,5} - 4_{2,2}$ transitions at 183.3, 321.2, and 325.2~GHz, respectively. We also examine their potential as probes of evolutionary stage by comparing them with previously reported Class~I methanol masers (MM).}
     {We used the SEPIA~180 and 345 receivers on the APEX 12\,m telescope to carry out the observations. To assess the evolutionary stage of each source, we modeled their spectral energy distributions (SEDs) using archival data and used the derived dust temperatures as proxies of ages. We then compared the occurrence of water and methanol masers across the sample.}
     {We detected 183.3~GHz water masers in 5 out of 18 sources. IRAS~16293--2422 shows all three transitions, while Serpens~FIRS~1 also displays the 321.2~GHz line. Despite excellent observing conditions, detection rates drop with increasing frequency, reflecting both intrinsic line weakness and variability. Notably, the brightest (sub)millimeter masers can reach flux densities comparable to the 22~GHz line. Comparisons of velocity profiles show that different transitions often trace distinct gas components.
     Water masers generally appear at earlier or comparable evolutionary stages than MM, suggesting no universal maser-based age sequence.   }
  {Our results demonstrate the detectability of submillimeter water in low-mass SFRs, although their occurrence is sparse. Velocity overlap between some centimeter and millimeter components suggests partial spatial coincidence, but many features appear uniquely in one frequency regime, indicating that different transitions often trace distinct gas regions with varying physical conditions. 
  }

   
\titlerunning{Survey of water masers in low-mass star-forming regions}
\authorrunning{Humire et al. (2025)}

   \maketitle
%

\section{Introduction}
\label{sec.intro}

Water masers are widely observed in star-forming regions (SFRs), where they trace shocked gas associated with collimated jets, disk winds, and outflows driven during the earliest stages of protostellar evolution \citep[e.g.,][]{Goddi2005, Greenhill2013, Moscadelli2019}. In particular, the $J=6_{1,6} - 5_{2,3}$ rotational transition of ortho-H$_2$O at 22.235~GHz has been detected in hundreds of sources, both in high- and low-mass SFRs \citep{Furuya2003, Moscadelli2006, Moscadelli2016}. Owing to their exceptional brightness and compactness, 22~GHz water masers are ideal targets for very long baseline interferometry (VLBI) observations, which have been critical for probing the dense gas and dynamical processes around young stellar objects (YSOs) \citep[e.g.,][]{Goddi2006,Moscadelli2020}.

Other masing transitions of H$_2$O occur in the (sub)millimeter (mm) and far-infrared (FIR) bands \citep{Neufeld1991}, but are less studied largely due to the observational challenges posed by atmospheric absorption, especially near 183 and 325~GHz \citep[see, e.g.,][]{Peck2018}. 
In high-mass SFRs, water maser emission has been detected at 183, 232, 321, 325, 439, 471, and 658~GHz \citep{Humphreys2007, Hirota2012, Hirota2016}, while para-H$_2$O transitions at 183 and 325~GHz have also been observed in low-mass SFRs \citep{Menten1990, Humphreys2007}. To date, (sub)mm water masers have been identified in only a handful of sources, and the physical conditions required for their excitation remain poorly constrained.
The 22 and 321~GHz ortho-H$_2$O masers are believed to originate in regions with distinct physical properties \citep{Neufeld1990, Neufeld1991}. In particular, the 321~GHz transition is thought to arise in warmer environments than the 22~GHz masers \citep{Patel2007}. Conversely, the 325~GHz emission appears to trace material across a broad range of densities, including low- and high-density regions \citep{Cernicharo1999, Niederhofer2012}. 
In Orion BN/KL, the 183~GHz and 325~GHz maser emissions are more extended than the 22~GHz emission, with 183~GHz showing the widest distribution \citep{Cernicharo1999}. This is consistent with their excitation requirements: the 183~GHz line (upper energy level at 205~K) can arise in cooler, less dense gas than the 22~GHz line (644~K), while the 325 GHz transition (470~K) lies in between.

Most previous searches for (sub)mm water masers have focused on high-mass SFRs, including well-known regions such as Orion, W49N, Cep A, W3(OH), W51, and Sgr~B2 \citep[e.g.,][]{Humphreys2007, Wang2023}. In contrast, only a limited number of studies have targeted low-mass SFRs, with observations toward sources such as Orion, HH,7–11, L1448IRS3, L1448–mm, IRAS 16293–2422, and Serpens-SMM1 \citep{Menten1990, Cernicharo1990, Cernicharo1996, vanKempen2009, Kang2013}. These studies have shown that the flux densities of the 183~GHz and 325~GHz maser transitions are comparable to those observed in the 22~GHz line.

In this study, we present a systematic search for submm water masers in low-mass SFRs. We compiled a comprehensive list of objects known to host 22~GHz water maser emission and selected 18 sources with declinations $\delta < 20^{\circ}$ \citep[][]{Furuya2003, Kang2013, deGM2006, Wilking1994, Healy2004}. 
Among them, we include four newly discovered 22~GHz maser sources associated with Class~0 and Class~I protostars in the Serpens South cluster \citep{Ortiz-Leon2021}, a region of active low-mass star formation \citep{Gutermuth2008}. This work builds on our previous study of the same sample, which focused on methanol maser emission \citep{Humire2024}. Here, we aim to investigate whether submm water masers are also present in these sources, thereby providing new constraints on the physical conditions and excitation mechanisms in low-mass protostellar environments.

The structure of this paper is as follows. In Section~\ref{Sec.Observations}, we describe the observational setup and strategy used to search for water maser emission in our sample. Section~\ref{Sec.Results} presents the detections and compares the velocity profiles of the newly observed maser lines with those of the 22.2GHz transition. In Section~\ref{Sec.discussion}, we discuss the diagnostic value of multi-frequency water masers and explore their potential role in an evolutionary framework, by comparing our results with previously reported methanol masers in the same sources \citep{Humire2024}. Finally, we summarize our main findings in Section~\ref{sec.conclusions}.

\section{Observations}
\label{Sec.Observations}

\subsection{The sample}
\label{subsec.subsample}

We conducted a survey targeting 18 low-mass SFRs, as listed in Table~\ref{tab:H2O_detections_coords_vlsr}. All selected sources exhibit previously detected maser emission in the $6_{1,6} - 5_{2,3}$ transition of H$_2$O at 22.235\,GHz. Our primary objective is to search for (sub)millimeter water maser emission in three higher-frequency transitions: $3_{1,3} - 2_{2,0}$ at 183.308\,GHz, $10_{2,9} - 9_{3,6}$ at 321.226\,GHz, and $5_{1,5} - 4_{2,2}$ at 325.153\,GHz.
The targets were selected from the original sample of 20 sources presented in \citet{Humire2024}. Two sources from that list were excluded from the present study: VLA\,1623 was not observed due to telescope scheduling constraints, and CARMA\,6 could not be spatially resolved from CARMA\,7 at 183\,GHz due to limited angular resolution (although they are distinguishable at 325\,GHz). As a result, our final sample consists of 18 targets. For a detailed description of the individual sources, we refer the reader to \citet{Humire2024} and Appendix~\ref{sec:sources}.

\subsection{ 
Observations and data reduction}
\label{Sec.observations}

Observations were conducted with the APEX 12\,m telescope at Llano de Chajnantor, Chile \citep{Guesten2006}, during three runs in April–June 2022 (projects M-0109.F-9512B/C-2022, P.I. A. Hern\'andez-G\'omez). We used the SEPIA~180 and SEPIA~345 receivers \citep{Belistsky2018, Meledin2022}. The FFTS4G backend provided 8\,GHz total bandwidth, and observations were performed in wobbler-switching mode with a 120$''$ throw. The observed bands covered 181.140–185.140\,GHz (USB) and 193.480–197.480\,GHz (LSB) with SEPIA~180 and 323.153–327.153\,GHz (USB) and 335.153–339.153\,GHz (LSB) with SEPIA~345. The spectral resolution was resampled to a common 0.1\,km\,s$^{-1}$ channel width. 

We measure a $3\sigma$ sensitivity of approximately 0.1~K or 3~Jy, calculated as the average of the median sensitivities from featureless spectral regions in the four bands mentioned above. The half-power beam width (HPBW) is 34$\arcsec$ at 181~GHz and 19.2$\arcsec$ at 325~GHz. Given the telescope's location, we benefit from excellent observing conditions. Specifically for the five sources highlighted in this work (see Appendix~\ref{sec:sources}), the median precipitable water vapor level is of 0.546~mm.

The delivered antenna temperature $T_{\rm{A}}^*$ was converted to main beam brightness temperature $T_{\rm{MB}}$ using $T_{\rm{MB}} = T_{\rm{A}}^* (\eta_{\rm{fw}} / \eta_{\rm{MB}})$, where the forward efficiency is $\eta_{\rm{fw}} = 0.95$ and the main-beam efficiency is computed as $\eta_{\rm{MB}} = 1.2182 \times \eta_{\rm{a}}$, with $\eta_{\rm{a}} = 0.71$ for SEPIA~180 and $\eta_{\rm{a}} = 0.67$ for SEPIA~345\footnote{Efficiency values from \url{http://www.apex-telescope.org/telescope/efficiency/index.php}} (measured from observations toward Mars). This results in $\eta_{\rm{MB}} = 0.865$ for SEPIA~180 and $\eta_{\rm{MB}} = 0.816$ for SEPIA~345. The absolute flux-scale calibration uncertainty is estimated to be $\sim$10\% \citep{Dumke2010}.
Data reduction was performed using the IRAM GILDAS/CLASS package.
Further observational details such as system temperatures, observing times, pointing and focus calibrators, and smoothing techniques are provided in \citet{Humire2024}.

\section{Results}
\label{Sec.Results}

\subsection{Water maser detections}

To search for the targeted water maser transitions in our sample, we employed the Centre d’Analyse Scientifique de Spectres Instrumentaux et Synthétiques (CASSIS; \citealt{Vastel2015}) as a front-end to the Cologne Database for Molecular Spectroscopy (CDMS; \citealt{Mueller2005}) and the Jet Propulsion Laboratory (JPL; \citealt{Pickett1998}) catalogs.

\begin{figure*}[!ht]
\centering
\includegraphics[width=0.99\textwidth, trim={0 1.5cm 0 2.5cm}, clip]{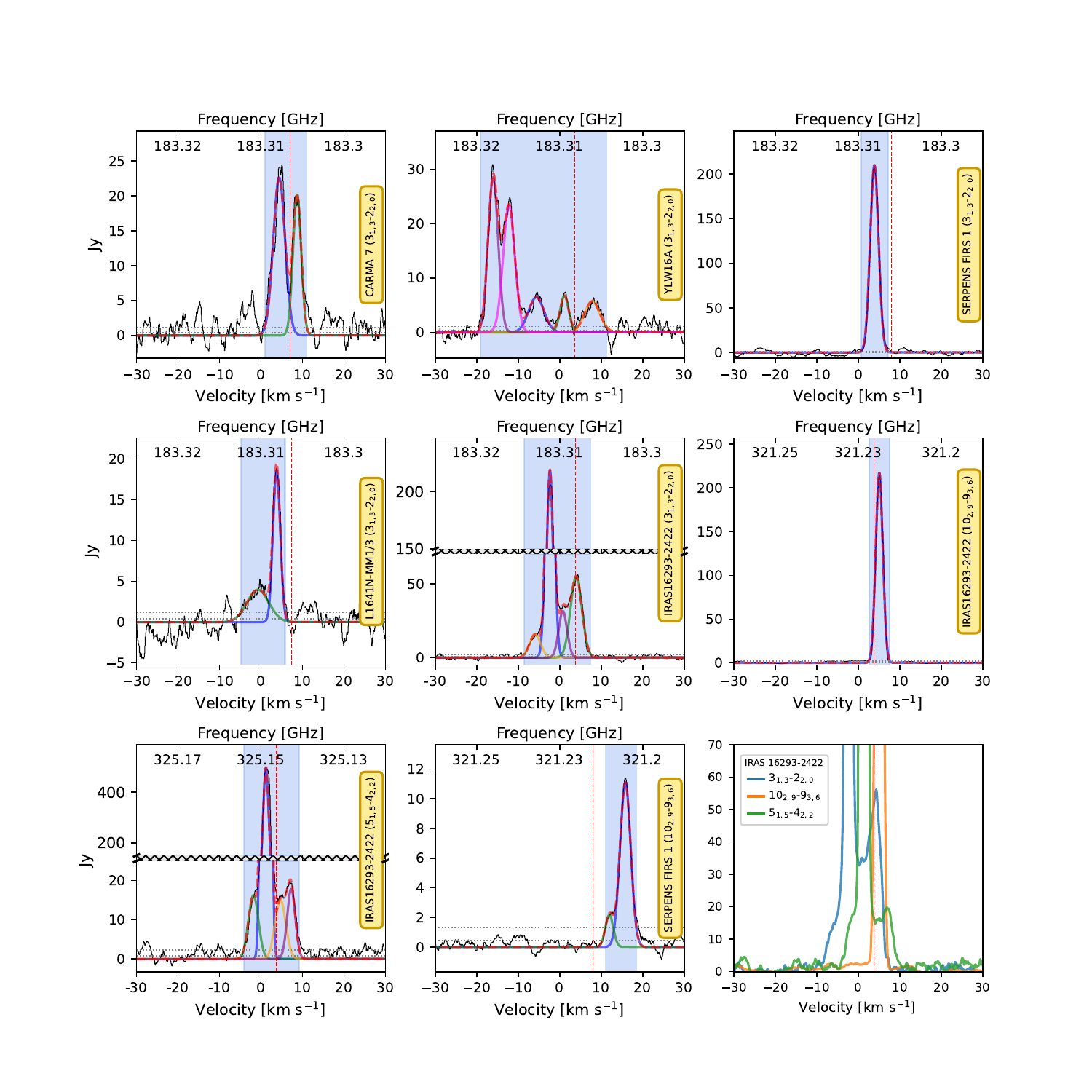}
\caption{Water maser transitions detected in our sample (Table~\ref{tab:H2O_detections_coords_vlsr}). Source names and maser transitions are indicated to the right of each sub-panel. Velocities are given in the observed frame. The observed spectra are displayed in black, while single or multiple Gaussian fits are overplotted as colored, semi-transparent solid lines. The total fitted emission is shown as a dashed red line. Vertical dashed red lines indicate the systemic velocities of the sources, as listed in Table~\ref{tab:H2O_detections_coords_vlsr}. For a few panels, the ordinate axis is limited between 20 and 150~Jy to better highlight weaker features. Shaded blue regions indicate the full velocity range of the maser emission, defined as the interval where the fitted model exceeds 3$\sigma$ above the baseline noise. Horizontal dashed black lines mark the 1 and 3$\sigma$ noise levels, estimated over the shown velocity range excluding the shaded regions. The bottom right panel combines all IRAS~16293--2422 sub(mm) transitions with a flux limit of 70~Jy to highlight weak spectral features.
}
\label{fig:profiles}
\end{figure*}

Five sources in the sample exhibit emission from the $3_{1,\,3} - 2_{2,\,0}$ transition at 183~GHz, as summarized in Table~\ref{tab.proposed_evol_sequence}. 
The 321~GHz transition is detected in only two sources: IRAS~16293--2422 and Serpens~FIRS~1. The 325~GHz maser line is detected solely in IRAS~16293--2422, where it exhibits a peak flux density of 497.4~Jy.
The corresponding spectral profiles are displayed in Figure~\ref{fig:profiles}. 
With the exception of the $5_{1,5} - 4_{2,2}$ emission in IRAS~16293--2422 \citep{Menten1990} and the $3_{1,3} - 2_{2,0}$ line in Serpens~FIRS~1 \citep{vanKempen2009}, all other detections presented here are reported for the first time.

\subsection{Millimeter transitions at 183, 321, and 325~GHz}

The measured velocities and flux densities of individual maser components, obtained via multi-Gaussian fitting to the spectral line profiles, are presented in Table~\ref{tab.compdetections}.
Among the three observed millimeter transitions, the 183~GHz line is the most frequently detected, appearing in all five sources with two to four distinct velocity components each. Peak flux densities span a wide range, from a few Jy up to over 200~Jy. The most prominent detections are toward IRAS~16293--2422 and Serpens~FIRS~1. Remarkably, IRAS~16293--2422 is the only source among the 18 observed that exhibits emission from both the $10_{2,\,9} - 9_{3,\,6}$ and $5_{1,\,5} - 4_{2,\,2}$ H$_2$O maser transitions at 321 and 325~GHz, respectively. The 325~GHz transition shows the highest peak intensity, reaching approximately 497~Jy. Serpens~FIRS~1 also shows a detection of the $10_{2,\,9} - 9_{3,\,6}$ maser line, which has an upper-state energy of $E_{\rm{up}} = 1861$~K, possibly indicative of stronger shocks or elevated excitation conditions in these two sources relative to the rest of the sample.

In terms of velocity structure, the 183 and 325~GHz components are generally found within $\sim$5--10~km\,s$^{-1}$ of the systemic velocities of their respective sources, though broader deviations are also observed. For example, a 183~GHz component in YLW16A is found at $-16.0$~km\,s$^{-1}$, substantially blueshifted relative to the systemic velocity of 3.6~km\,s$^{-1}$. The 321~GHz transition in Serpens~FIRS~1 displays primarily redshifted components, reaching up to $\sim$16~km\,s$^{-1}$, while its 183~GHz component is blueshifted relative to the systemic value of 8.0~km\,s$^{-1}$. 

Overall, the velocity spreads of these maser lines typically range from 10 to 20~km\,s$^{-1}$, with a limited number of discrete components per source.

\subsection{Comparison between mm and cm emission}
\label{cm-mm_comp}
The properties of the 22~GHz maser emission, shown in the final two columns of Table~\ref{tab.compdetections}, excluding the reference column, are compiled from previous studies conducted at different epochs and resolutions. Specifically, we adopt data from CARMA~7 \citep{Ortiz-Leon2021}, Serpens~FIRS~1 \citep{Furuya2003, Bae2011}, L1641N\,MM1/3 \citep{Bae2011, Kang2013}, YLW16A \citep{Furuya2003, Sunada2007}, and IRAS~16293--2422 \citep{Furuya2003, Sunada2007}. When multiple nearby velocity components are reported, we bin them in 0.5~km\,s$^{-1}$ intervals and compute the average velocity. For sources observed across multiple epochs or studies, we adopt the average peak flux density.

A few comparative trends can be noted between the millimeter and centimeter water maser emission. 
The 22~GHz masers consistently exhibit a larger number of velocity components than any of the millimeter transitions. For instance, IRAS~16293--2422 shows at least nine distinct components at 22~GHz, in contrast to two to four components typically observed at 183, 321, or 325~GHz. Flux densities at 22~GHz are comparable to or exceed those of the brightest millimeter masers. While the 325~GHz line in IRAS~16293--2422 reaches nearly 500~Jy, the 22~GHz maser line shows similarly high peak fluxes across several sources, including 366~Jy in YLW16A.

In terms of velocity coverage, the 22~GHz masers generally span a broader range than their mm counterparts. For example, in YLW16A, 22~GHz components extend from $-15.2$ to $+16.3$~km\,s$^{-1}$, whereas mm transitions in the same source exhibit a narrower spread centered closer to the systemic velocity.
There are several instances where velocity components from the mm and cm lines coincide within a few km\,s$^{-1}$. However, many features appear in only one frequency regime, suggesting that these transitions trace physically distinct masing regions.

A final caveat warrants emphasis: this comparison is  constrained by the non-simultaneity of the observations, as maser emission is known to exhibit significant time variability. Moreover, the identification of velocity-coincident components does not imply spatial coincidence; confirming physical associations among different maser transitions requires spatially resolved, multi-frequency observations conducted contemporaneously.

\section{Discussion}
\label{Sec.discussion}

\subsection{cm and mm H$_2$O maser lines as probes of star formation}

The 22~GHz H$_2$O maser is a well-established tracer of shocked gas at the interface between protostellar jets and the surrounding medium, widely used to probe outflows on scales of tens to hundreds of astronomical units (AU) \citep{Goddi2017,Moscadelli2019}.

This work presents a first attempt to detect (sub)millimeter water masers at 183, 321, and 325~GHz in low-mass YSOs. These transitions, known in high-mass SFRs, are predicted to be strongly inverted under conditions similar to the 22~GHz line ($n_{\rm H_2} \sim 10^7$–$10^{10}$ cm$^{-3}$, $T_{\rm kin} \sim 400$–3000~K; \citealt{Gray2016}).

So far, mm/submm water masers have been imaged in just two high-mass regions (Orion-KL and Cep~A) and one low-mass source (Serpens SMM1). In Orion, the 22, 321, and 325~GHz masers trace the same bipolar outflow, suggesting similar excitation conditions \citep{Niederhofer2012, Greenhill2013}. In Cep~A, however, 321~GHz masers appear to trace hotter inner regions not seen at 22~GHz \citep{Patel2007}.

Our comparison (Section~\ref{cm-mm_comp}) finds several mm and cm maser components overlapping in velocity, but many occur independently. This suggests that the transitions often probe different gas regions with varying physical conditions such as temperature, density, or shock velocity.
If co-spatial, multi-line observations allow detailed modeling of gas conditions. If not, they offer a layered view of the circumstellar structure. High-resolution interferometry is essential to distinguish between these cases.
In both scenarios, combining cm and mm H$_2$O maser lines offers powerful diagnostics of shocked gas in star formation—either by constraining physical conditions where lines overlap, or by mapping distinct regions in protostellar environments.
\subsection{An evolutionary scenario?}
\label{sec.evol_scenario}

Several studies have proposed that masers trace the early evolutionary stages of YSOs, with different molecular species and transitions appearing at different times \citep[e.g.,][]{Ellingsen2007, Reid2007, Urquhart2024}. In this context, masers could potentially serve as diagnostics for the protostellar phase.

Methanol masers (MMs) are traditionally divided into two classes: Class~I, collisionally pumped and typically found in outflows about 1~pc from ultra-compact H~\textsc{ii} regions, and Class~II, radiatively pumped and tightly associated with high-mass YSOs \citep{Menten1991b,Billington2020}. Notably, both classes can appear before the onset of detectable continuum emission \citep{Ellingsen1996a, Zeng2020}, implying that they trace early phases of star formation. 
Class II MM are also observed along with H$_2$O masers \citep{Codella2004, Goddi2011}, which are commonly associated with shocks in protostellar jets and outflows.
However, establishing a strict evolutionary sequence based on maser species remains observationally challenging \citep[e.g.,][]{Voronkov2014}. In high-mass YSOs, OH masers are thought to trace more evolved stages, while Class~I MM  (e.g., at 84 and 95~GHz) appear earlier \citep{Yang2023}. Unfortunately, OH masers are absent in low-mass YSOs, limiting their diagnostic power in our context.

\begin{table}[h!]
\setlength{\tabcolsep}{0.5pt} 
\centering
\tiny
\caption{Proposed evolutionary scenario.}
\begin{tabular}{|l|c|c|c|c|c|c|c|c|}
\hline \hline
Region & \multicolumn{2}{c|}{MMcIs$J_{-1}$} & \multicolumn{4}{c|}{H$_2$O} & $T_{\rm{DD}}$ [K] & Relative age\\ \hline
\textbf{} & J=7 & J=10 & 22 & 183 & 321 & 325 & \textbf{} & \\  
\textbf{} & 70.6 & 133.2 & 643.5 & 204.71 & 1861.3 & 469.9 & &  \\ \hline
IRAS 16293  & \XSolidBrush & \XSolidBrush & \Checkmark & \Checkmark & \Checkmark & \Checkmark & 16.92$_{-4.70}^{+10.20}$ & \text{quite young} \\
CARMA~7         & \Checkmark & \Checkmark & \Checkmark & \Checkmark & \XSolidBrush & \XSolidBrush & 29.56$_{-2.50}^{+2.89}$ & \text{very young} \\
YLW16A          & \XSolidBrush & \XSolidBrush & \Checkmark & \Checkmark & \XSolidBrush & \XSolidBrush & 37.93$_{-11.14}^{+28.00}$ & \text{young} \\
Serpens~FIRS    & \XSolidBrush & \Checkmark & \Checkmark & \Checkmark & \Checkmark & \XSolidBrush & 42.07$_{-17.61}^{+15.42}$ & \text{slightly evolved} \\
L1641N          & \Checkmark & \Checkmark & \Checkmark & \Checkmark & \XSolidBrush & \XSolidBrush & 48.493$_{-11.54}^{+6.18}$ & \text{evolved} \\
NGC~2024        & \XSolidBrush & \Checkmark & \Checkmark & \XSolidBrush & \XSolidBrush & \XSolidBrush & 94.04$_{-44.80}^{+62.58}$ & \text{very evolved} \\
\hline
\end{tabular}
\tablefoot{
Frequencies of the water masers are indicated in GHz, with upper energy levels in Kelvin shown below each transition label. The diffuse dust temperatures ($T_{\rm{DD}}$) were derived from SED fits (see Appendix~\ref{appen.SED_modeling}).
} 
\label{tab.proposed_evol_sequence}
\end{table}

\begin{figure}[!ht]
\centering
\includegraphics[width=0.51\textwidth, trim={0.4cm 0.55cm 0 0.3cm}, clip]{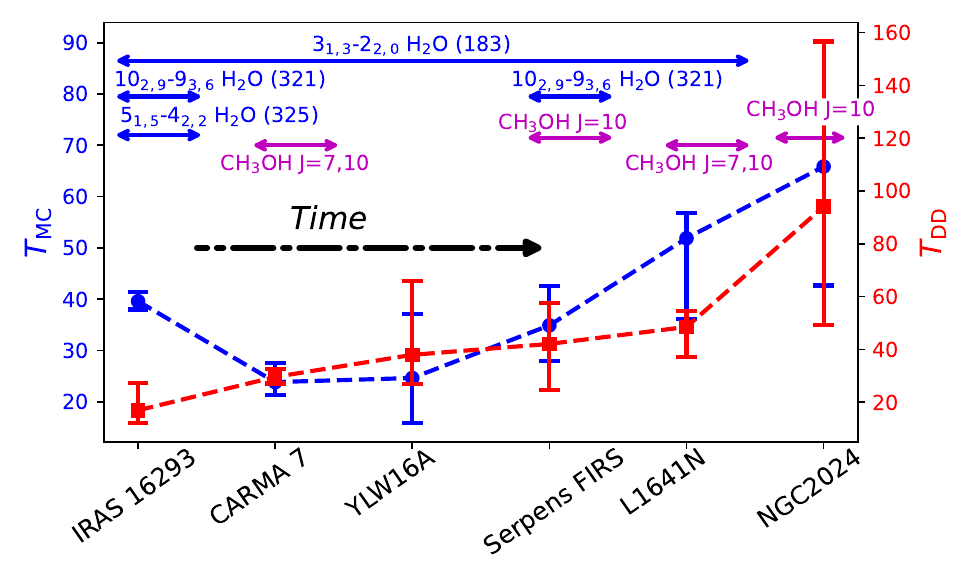}
\caption{
Sources are ordered along the x-axis from youngest to oldest, based on the diffuse dust temperature $T_{\rm{DD}}$ (see Table~\ref{tab.proposed_evol_sequence}), which is shown in red and corresponds to the rightmost y-axis. The left y-axis shows the molecular cloud temperature ($T_{\rm{MC}}$, in blue), derived from graybody fits to the SED using the same methodology as for $T_{\rm{DD}}$ (see Appendix~\ref{appen.SED_modeling}). Water and methanol maser detections are indicated above each source, with water maser transitions labeled by their frequency in GHz. All temperatures are in Kelvin.
}
\label{fig:evol_masers}
\end{figure}

In a recent study, \citet{Humire2024} detected MM in the $J_{-1}\rightarrow(J-1)_0$-E series for the first time in low-mass YSOs, at $J=7$ and 10 (181 and 325~GHz). We refer to these as MMcIs7 and MMcIs10, respectively. If the maser-based evolutionary framework developed \cg{proposed} for massive YSOs \citep[e.g.,][]{Ellingsen2007} also applies to low-mass objects, then sources exhibiting both 22~GHz water and MMcIs masers should be younger than those with only water masers.
To test this scenario, we compare the presence of H$_2$O and CH$_3$OH masers across our sample using an independent age indicator: the diffuse dust temperature ($T_{\text{DD}}$), derived from SED modeling (see Appendix~\ref{appen.SED_modeling} and Fig.~\ref{Fig.SEDs}). We choose $T_{\text{DD}}$ over traditional chemical clocks (e.g., S-bearing molecules like SO or OCS; see \citealt{Wakelam2004, Herpin2009}) due to known limitations, which are further expanded upon in Appendix~\ref{Appendix.sulfur}.

Figure~\ref{fig:evol_masers} shows the distribution of maser detections as a function of $T_{\text{DD}}$. This plot suggests that water masers tend to appear earlier than methanol masers in our sample of low-mass YSOs. While this trend appears to contrast with the evolutionary scenario proposed by \citet{Ellingsen2007} for high-mass YSOs, 
several caveats must be considered.
(1) Our analysis is based solely on water and methanol masers, whereas \citet{Ellingsen2007} also includes OH masers, which are not typically detected in low-mass YSOs. This limits the basis for a direct comparison.
(2) The evolutionary paths of low-mass and high-mass YSOs may differ significantly, implying that maser-based diagnostics might follow different trends.
(3) Our sample is relatively small and lacks the statistical robustness needed for firm conclusions.

With these limitations in mind, we nonetheless explore the possibility that, in low-mass protostars, water masers may indeed emerge earlier than methanol masers in the course of their evolution.
Water masers are more robust under high-velocity shock conditions ($v > 10$~km~s$^{-1}$), where methanol molecules are more easily destroyed. H$_2$O, in contrast, can reform efficiently in post-shock gas \citep{Suutarinen2014}, and has higher optical depth in C-type shocks at $v > 15$~km~s$^{-1}$ \citep{Nesterenok2022}. These factors may make water masers more likely to appear first, especially during the earliest, most dynamically active phases of low-mass star formation.
Additionally, our data allow comparison across multiple water maser transitions, from the canonical 22~GHz (cm) to higher-frequency (mm) lines at 183, 321, and 325~GHz. While the 22~GHz line typically traces compact and bright shocked regions, the mm transitions can trace more extended structures (e.g., 183~GHz) or warmer gas (e.g., 321~GHz with $E_{\rm up}\sim$1800~K). However, we find no clear trend linking specific H$_2$O transitions to distinct evolutionary phases—possibly due to the overlap of excitation conditions and variability timescales.

In summary, our results suggest that the presence or absence of individual maser transitions cannot serve as a reliable age diagnostic in low-mass YSOs. Water masers, particularly at 22~GHz, appear to be more ubiquitous and may precede MM in these environments. This challenges simple maser-based evolutionary schemes and highlights the need for multi-tracer, multi-epoch observations to robustly assess YSO evolution.

\section{Conclusions}
\label{sec.conclusions}

We conducted a survey of H$_2$O masers in 18 low-mass SFRs, all previously known to exhibit emission in the 22~GHz ($6_{1,6}-5_{2,3}$) transition. Our goal was to explore the presence of additional water maser transitions at 183, 321, and 325~GHz, to study their excitation conditions, and to assess their diagnostic potential in relation to evolutionary stages of YSOs.

Our main findings are as follows:

\begin{itemize}
    \item We detected the 183~GHz ($3_{1,3}-2_{2,0}$) water maser emission in five sources: IRAS~16293--2422, CARMA~7, Serpens~FIRS~1, YLW16A, and L1641N-MM1/3.  Among these, IRAS~16293--2422 is the only source that also shows maser emission in both the 321~GHz ($10_{2,9}-9_{3,6}$) and 325~GHz ($5_{1,5}-4_{2,2}$) transitions.
    \item By combining these maser data with  SED modeling, we used dust temperature as a proxy for evolutionary stage. We find that water masers often appear before or concurrently with Class~I MM, suggesting that masing phenomena using only these two molecules can not be used as a reliable age indicator for low-mass YSOs.
    \item We find that several cm (from the literature) and mm (from this work) maser components occur at similar velocities, but many features are unique to a specific transition. This indicates physically distinct gas regions with differing densities, temperatures, or kinematics between them.
    \item While 22~GHz water masers are well-established probes of shocked gas at jet–ambient interfaces, additional detections at 183–325~GHz offer complementary diagnostics. If the maser lines trace the same gas, multi-frequency modeling can constrain physical conditions. If not, they provide a layered view of the circumstellar environment at different spatial scales or excitation regimes.

\end{itemize}

In summary, multi-frequency water masers—spanning cm to submm wavelengths—offer a promising but complex toolset for probing the early stages of low-mass star formation. High-resolution interferometric imaging of submm lines will be essential to fully exploit their diagnostic potential, clarify the origin of masing regions, and test excitation models.

\begin{acknowledgements}
We thank the anonymous referee for their helpful comments, questions, and suggestions on revising the manuscript. P.K.H. gratefully acknowledges the Fundação de Amparo à Pesquisa do Estado de São Paulo (FAPESP) for the support grant 2023/14272-4. 
C.G. acknowledges financial support by FAPESP under grant 2021/01183-8 and the European Union NextGenerationEU RRF M4C2 1.1 project n. 2022YAPMJH. G.N.O.L. acknowledges the financial support provided by the Instituto Nacional de Astrof{\'i}sica, \'Optica y Electr{\'o}nica, and Secretar{\'i}a de Ciencia, Humaninadades, Tecnolog{\'i}a e Innovaci{\'o}n.

\end{acknowledgements}

\hypertarget{bib:Humire2024}{}
\bibliographystyle{aa}
\bibliography{aanda} 

\begin{thebibliography}{80}
\expandafter\ifx\csname natexlab\endcsname\relax\def\natexlab#1{#1}\fi

\bibitem[{{Allen} {et~al.}(2002){Allen}, {Myers}, {Di Francesco}, {Mathieu}, {Chen}, \& {Young}}]{Allen2002}
{Allen}, L.~E., {Myers}, P.~C., {Di Francesco}, J., {et~al.} 2002, \apj, 566, 993

\bibitem[{{Artur de la Villarmois} {et~al.}(2019){Artur de la Villarmois}, {J{\o}rgensen}, {Kristensen}, {Bergin}, {Harsono}, {Sakai}, {van Dishoeck}, \& {Yamamoto}}]{Artur_de_la_Villarmois2019}
{Artur de la Villarmois}, E., {J{\o}rgensen}, J.~K., {Kristensen}, L.~E., {et~al.} 2019, \aap, 626, A71

\bibitem[{{Bae} {et~al.}(2011){Bae}, {Kim}, {Youn}, {Kim}, {Byun}, {Kang}, \& {Oh}}]{Bae2011}
{Bae}, J.-H., {Kim}, K.-T., {Youn}, S.-Y., {et~al.} 2011, \apjs, 196, 21

\bibitem[{{Belitsky} {et~al.}(2018){Belitsky}, {Lapkin}, {Fredrixon}, {Meledin}, {Sundin}, {Billade}, {Ferm}, {Pavolotsky}, {Rashid}, {Strandberg}, {Desmaris}, {Ermakov}, {Krause}, {Olberg}, {Aghdam}, {Shafiee}, {Bergman}, {De Beck}, {Olofsson}, {Conway}, {De Breuck}, {Immer}, {Yagoubov}, {Montenegro-Montes}, {Torstensson}, {P{\'e}rez-Beaupuits}, {Klein}, {Boland}, {Baryshev}, {Hesper}, {Barkhof}, {Adema}, {Bekema}, \& {Koops}}]{Belistsky2018}
{Belitsky}, V., {Lapkin}, I., {Fredrixon}, M., {et~al.} 2018, \aap, 612, A23

\bibitem[{{Billington} {et~al.}(2020){Billington}, {Urquhart}, {K{\"o}nig}, {Beuther}, {Breen}, {Menten}, {Campbell-White}, {Ellingsen}, {Thompson}, {Moore}, {Eden}, {Kim}, \& {Leurini}}]{Billington2020}
{Billington}, S.~J., {Urquhart}, J.~S., {K{\"o}nig}, C., {et~al.} 2020, \mnras, 499, 2744

\bibitem[{{Cernicharo} {et~al.}(1996){Cernicharo}, {Bachiller}, \& {Gonzalez-Alfonso}}]{Cernicharo1996}
{Cernicharo}, J., {Bachiller}, R., \& {Gonzalez-Alfonso}, E. 1996, \aap, 305, L5

\bibitem[{{Cernicharo} {et~al.}(1999){Cernicharo}, {Pardo}, {Gonz{\'a}lez-Alfonso}, {Serabyn}, {Phillips}, {Benford}, \& {Mehringer}}]{Cernicharo1999}
{Cernicharo}, J., {Pardo}, J.~R., {Gonz{\'a}lez-Alfonso}, E., {et~al.} 1999, \apjl, 520, L131

\bibitem[{{Cernicharo} {et~al.}(1990){Cernicharo}, {Thum}, {Hein}, {John}, {Garcia}, \& {Mattioco}}]{Cernicharo1990}
{Cernicharo}, J., {Thum}, C., {Hein}, H., {et~al.} 1990, \aap, 231, L15

\bibitem[{{Chabrier}(2003)}]{Chabrier2003}
{Chabrier}, G. 2003, \pasp, 115, 763

\bibitem[{{Choi} {et~al.}(2012){Choi}, {Kang}, {Byun}, \& {Lee}}]{Choi2012}
{Choi}, M., {Kang}, M., {Byun}, D.-Y., \& {Lee}, J.-E. 2012, \apj, 759, 136

\bibitem[{{Codella} {et~al.}(2004){Codella}, {Lorenzani}, {Gallego}, {Cesaroni}, \& {Moscadelli}}]{Codella2004}
{Codella}, C., {Lorenzani}, A., {Gallego}, A.~T., {Cesaroni}, R., \& {Moscadelli}, L. 2004, \aap, 417, 615

\bibitem[{{Crimier} {et~al.}(2010){Crimier}, {Ceccarelli}, {Maret}, {Bottinelli}, {Caux}, {Kahane}, {Lis}, \& {Olofsson}}]{Crimier2010}
{Crimier}, N., {Ceccarelli}, C., {Maret}, S., {et~al.} 2010, \aap, 519, A65

\bibitem[{{Curiel} {et~al.}(1996){Curiel}, {Rodriguez}, {Gomez}, {Torrelles}, {Ho}, \& {Eiroa}}]{Curiel1996}
{Curiel}, S., {Rodriguez}, L.~F., {Gomez}, J.~F., {et~al.} 1996, \apj, 456, 677

\bibitem[{{de Gregorio-Monsalvo} {et~al.}(2006){de Gregorio-Monsalvo}, {G{\'o}mez}, {Su{\'a}rez}, {Kuiper}, {Rodr{\'\i}guez}, \& {Jim{\'e}nez-Bail{\'o}n}}]{deGM2006}
{de Gregorio-Monsalvo}, I., {G{\'o}mez}, J.~F., {Su{\'a}rez}, O., {et~al.} 2006, \apj, 642, 319

\bibitem[{{Drozdovskaya} {et~al.}(2018){Drozdovskaya}, {van Dishoeck}, {J{\o}rgensen}, {Calmonte}, {van der Wiel}, {Coutens}, {Calcutt}, {M{\"u}ller}, {Bjerkeli}, {Persson}, {Wampfler}, \& {Altwegg}}]{Drozdovskaya2018}
{Drozdovskaya}, M.~N., {van Dishoeck}, E.~F., {J{\o}rgensen}, J.~K., {et~al.} 2018, \mnras, 476, 4949

\bibitem[{{Dumke} \& {Mac-Auliffe}(2010)}]{Dumke2010}
{Dumke}, M. \& {Mac-Auliffe}, F. 2010, in Society of Photo-Optical Instrumentation Engineers (SPIE) Conference Series, Vol. 7737, Observatory Operations: Strategies, Processes, and Systems III, ed. D.~R. {Silva}, A.~B. {Peck}, \& B.~T. {Soifer}, 77371J

\bibitem[{{Dzib} {et~al.}(2018){Dzib}, {Ortiz-Le{\'o}n}, {Hern{\'a}ndez-G{\'o}mez}, {Loinard}, {Mioduszewski}, {Claussen}, {Menten}, {Caux}, \& {Sanna}}]{Dzib2018}
{Dzib}, S.~A., {Ortiz-Le{\'o}n}, G.~N., {Hern{\'a}ndez-G{\'o}mez}, A., {et~al.} 2018, \aap, 614, A20

\bibitem[{{Ellingsen} {et~al.}(1996){Ellingsen}, {Norris}, \& {McCulloch}}]{Ellingsen1996a}
{Ellingsen}, S.~P., {Norris}, R.~P., \& {McCulloch}, P.~M. 1996, \mnras, 279, 101

\bibitem[{{Ellingsen} {et~al.}(2007){Ellingsen}, {Voronkov}, {Cragg}, {Sobolev}, {Breen}, \& {Godfrey}}]{Ellingsen2007}
{Ellingsen}, S.~P., {Voronkov}, M.~A., {Cragg}, D.~M., {et~al.} 2007, in Astrophysical Masers and their Environments, ed. J.~M. {Chapman} \& W.~A. {Baan}, Vol. 242, 213--217

\bibitem[{{Enoch} {et~al.}(2007){Enoch}, {Glenn}, {Evans}, {Sargent}, {Young}, \& {Huard}}]{Enoch2007}
{Enoch}, M.~L., {Glenn}, J., {Evans}, Neal~J., I., {et~al.} 2007, \apj, 666, 982

\bibitem[{{Froebrich}(2005)}]{Froebrich2005}
{Froebrich}, D. 2005, \apjs, 156, 169

\bibitem[{{Furuya} {et~al.}(2003){Furuya}, {Kitamura}, {Wootten}, {Claussen}, \& {Kawabe}}]{Furuya2003}
{Furuya}, R.~S., {Kitamura}, Y., {Wootten}, A., {Claussen}, M.~J., \& {Kawabe}, R. 2003, \apjs, 144, 71

\bibitem[{{Goddi} {et~al.}(2005){Goddi}, {Moscadelli}, {Alef}, {Tarchi}, {Brand}, \& {Pani}}]{Goddi2005}
{Goddi}, C., {Moscadelli}, L., {Alef}, W., {et~al.} 2005, \aap, 432, 161

\bibitem[{{Goddi} {et~al.}(2011){Goddi}, {Moscadelli}, \& {Sanna}}]{Goddi2011}
{Goddi}, C., {Moscadelli}, L., \& {Sanna}, A. 2011, \aap, 535, L8

\bibitem[{{Goddi} {et~al.}(2006){Goddi}, {Moscadelli}, {Torrelles}, {Uscanga}, \& {Cesaroni}}]{Goddi2006}
{Goddi}, C., {Moscadelli}, L., {Torrelles}, J.~M., {Uscanga}, L., \& {Cesaroni}, R. 2006, \aap, 447, L9

\bibitem[{{Goddi} {et~al.}(2017){Goddi}, {Surcis}, {Moscadelli}, {Imai}, {Vlemmings}, {van Langevelde}, \& {Sanna}}]{Goddi2017}
{Goddi}, C., {Surcis}, G., {Moscadelli}, L., {et~al.} 2017, \aap, 597, A43

\bibitem[{{Gray} {et~al.}(2016){Gray}, {Baudry}, {Richards}, {Humphreys}, {Sobolev}, \& {Yates}}]{Gray2016}
{Gray}, M.~D., {Baudry}, A., {Richards}, A.~M.~S., {et~al.} 2016, \mnras, 456, 374

\bibitem[{{Greenhill} {et~al.}(2013){Greenhill}, {Goddi}, {Chandler}, {Matthews}, \& {Humphreys}}]{Greenhill2013}
{Greenhill}, L.~J., {Goddi}, C., {Chandler}, C.~J., {Matthews}, L.~D., \& {Humphreys}, E.~M.~L. 2013, \apjl, 770, L32

\bibitem[{{G{\"u}sten} {et~al.}(2006){G{\"u}sten}, {Nyman}, {Schilke}, {Menten}, {Cesarsky}, \& {Booth}}]{Guesten2006}
{G{\"u}sten}, R., {Nyman}, L.~{\r{A}}., {Schilke}, P., {et~al.} 2006, \aap, 454, L13

\bibitem[{{Gutermuth} {et~al.}(2008){Gutermuth}, {Bourke}, {Allen}, {Myers}, {Megeath}, {Matthews}, {J{\o}rgensen}, {Di Francesco}, {Ward-Thompson}, {Huard}, {Brooke}, {Dunham}, {Cieza}, {Harvey}, \& {Chapman}}]{Gutermuth2008}
{Gutermuth}, R.~A., {Bourke}, T.~L., {Allen}, L.~E., {et~al.} 2008, \apjl, 673, L151

\bibitem[{{Healy} {et~al.}(2004){Healy}, {Hester}, \& {Claussen}}]{Healy2004}
{Healy}, K.~R., {Hester}, J.~J., \& {Claussen}, M.~J. 2004, \apj, 610, 835

\bibitem[{{Herpin} {et~al.}(2009){Herpin}, {Marseille}, {Wakelam}, {Bontemps}, \& {Lis}}]{Herpin2009}
{Herpin}, F., {Marseille}, M., {Wakelam}, V., {Bontemps}, S., \& {Lis}, D.~C. 2009, \aap, 504, 853

\bibitem[{{Hirota} {et~al.}(2012){Hirota}, {Kim}, \& {Honma}}]{Hirota2012}
{Hirota}, T., {Kim}, M.~K., \& {Honma}, M. 2012, \apjl, 757, L1

\bibitem[{{Hirota} {et~al.}(2016){Hirota}, {Kim}, \& {Honma}}]{Hirota2016}
{Hirota}, T., {Kim}, M.~K., \& {Honma}, M. 2016, \apj, 817, 168

\bibitem[{{Humire} {et~al.}(2024){Humire}, {Ortiz-Le{\'o}n}, {Hern{\'a}ndez-G{\'o}mez}, {Yang}, {Henkel}, \& {Mart{\'\i}n}}]{Humire2024}
{Humire}, P.~K., {Ortiz-Le{\'o}n}, G.~N., {Hern{\'a}ndez-G{\'o}mez}, A., {et~al.} 2024, \aap, 688, L1

\bibitem[{{Humphreys}(2007)}]{Humphreys2007}
{Humphreys}, E.~M.~L. 2007, in IAU Symposium, Vol. 242, Astrophysical Masers and their Environments, ed. J.~M. {Chapman} \& W.~A. {Baan}, 471--480

\bibitem[{{Jacobsen} {et~al.}(2018){Jacobsen}, {J{\o}rgensen}, {van der Wiel}, {Calcutt}, {Bourke}, {Brinch}, {Coutens}, {Drozdovskaya}, {Kristensen}, {M{\"u}ller}, \& {Wampfler}}]{Jacobsen2018}
{Jacobsen}, S.~K., {J{\o}rgensen}, J.~K., {van der Wiel}, M.~H.~D., {et~al.} 2018, \aap, 612, A72

\bibitem[{{Kang} {et~al.}(2013){Kang}, {Lee}, {Choi}, {Choi}, {Kim}, {Di Francesco}, \& {Park}}]{Kang2013}
{Kang}, M., {Lee}, J.-E., {Choi}, M., {et~al.} 2013, \apjs, 209, 25

\bibitem[{{Meledin} {et~al.}(2022){Meledin}, {Lapkin}, {Fredrixon}, {Sundin}, {Ferm}, {Pavolotsky}, {Strandberg}, {Desmaris}, {L{\'o}pez}, {Bergman}, {Olberg}, {Conway}, {Torstensson}, {Dur{\'a}n}, {Montenegro-Montes}, {De Breuck}, \& {Belitsky}}]{Meledin2022}
{Meledin}, D., {Lapkin}, I., {Fredrixon}, M., {et~al.} 2022, \aap, 668, A2

\bibitem[{{Menten}(1991)}]{Menten1991b}
{Menten}, K.~M. 1991, \apj, 380, L75

\bibitem[{{Menten} {et~al.}(1990){Menten}, {Melnick}, {Phillips}, \& {Neufeld}}]{Menten1990}
{Menten}, K.~M., {Melnick}, G.~J., {Phillips}, T.~G., \& {Neufeld}, D.~A. 1990, \apjl, 363, L27

\bibitem[{{Mezger} {et~al.}(1992){Mezger}, {Sievers}, {Haslam}, {Kreysa}, {Lemke}, {Mauersberger}, \& {Wilson}}]{Mezger1992}
{Mezger}, P.~G., {Sievers}, A.~W., {Haslam}, C.~G.~T., {et~al.} 1992, \aap, 256, 631

\bibitem[{{Mizuno} {et~al.}(1990){Mizuno}, {Fukui}, {Iwata}, {Nozawa}, \& {Takano}}]{Mizuno1990}
{Mizuno}, A., {Fukui}, Y., {Iwata}, T., {Nozawa}, S., \& {Takano}, T. 1990, \apj, 356, 184

\bibitem[{{Moscadelli} {et~al.}(2016){Moscadelli}, {S{\'a}nchez-Monge}, {Goddi}, {Li}, {Sanna}, {Cesaroni}, {Pestalozzi}, {Molinari}, \& {Reid}}]{Moscadelli2016}
{Moscadelli}, L., {S{\'a}nchez-Monge}, {\'A}., {Goddi}, C., {et~al.} 2016, \aap, 585, A71

\bibitem[{{Moscadelli} {et~al.}(2019){Moscadelli}, {Sanna}, {Goddi}, {Krishnan}, {Massi}, \& {Bacciotti}}]{Moscadelli2019}
{Moscadelli}, L., {Sanna}, A., {Goddi}, C., {et~al.} 2019, \aap, 631, A74

\bibitem[{{Moscadelli} {et~al.}(2020){Moscadelli}, {Sanna}, {Goddi}, {Krishnan}, {Massi}, \& {Bacciotti}}]{Moscadelli2020}
{Moscadelli}, L., {Sanna}, A., {Goddi}, C., {et~al.} 2020, \aap, 635, A118

\bibitem[{{Moscadelli} {et~al.}(2006){Moscadelli}, {Testi}, {Furuya}, {Goddi}, {Claussen}, {Kitamura}, \& {Wootten}}]{Moscadelli2006}
{Moscadelli}, L., {Testi}, L., {Furuya}, R.~S., {et~al.} 2006, \aap, 446, 985

\bibitem[{{M{\"u}ller} {et~al.}(2005){M{\"u}ller}, {Schl{\"o}der}, {Stutzki}, \& {Winnewisser}}]{Mueller2005}
{M{\"u}ller}, H. S.~P., {Schl{\"o}der}, F., {Stutzki}, J., \& {Winnewisser}, G. 2005, Journal of Molecular Structure, 742, 215

\bibitem[{{Nakamura} {et~al.}(2012){Nakamura}, {Miura}, {Kitamura}, {Shimajiri}, {Kawabe}, {Akashi}, {Ikeda}, {Tsukagoshi}, {Momose}, {Nishi}, \& {Li}}]{Nakamura2012}
{Nakamura}, F., {Miura}, T., {Kitamura}, Y., {et~al.} 2012, \apj, 746, 25

\bibitem[{{Nesterenok}(2022)}]{Nesterenok2022}
{Nesterenok}, A.~V. 2022, Astronomy Letters, 48, 345

\bibitem[{{Neufeld} \& {Melnick}(1990)}]{Neufeld1990}
{Neufeld}, D.~A. \& {Melnick}, G.~J. 1990, \apjl, 352, L9

\bibitem[{{Neufeld} \& {Melnick}(1991)}]{Neufeld1991}
{Neufeld}, D.~A. \& {Melnick}, G.~J. 1991, \apj, 368, 215

\bibitem[{{Niederhofer} {et~al.}(2012){Niederhofer}, {Humphreys}, {Goddi}, \& {Greenhill}}]{Niederhofer2012}
{Niederhofer}, F., {Humphreys}, E., {Goddi}, C., \& {Greenhill}, L.~J. 2012, in IAU Symposium, Vol. 287, Cosmic Masers - from OH to H0, ed. R.~S. {Booth}, W.~H.~T. {Vlemmings}, \& E.~M.~L. {Humphreys}, 184--185

\bibitem[{{Ortiz-Le{\'o}n} {et~al.}(2018){Ortiz-Le{\'o}n}, {Loinard}, {Dzib}, {Kounkel}, {Galli}, {Tobin}, {Evans}, {Hartmann}, {Rodr{\'\i}guez}, {Brice{\~n}o}, {Torres}, \& {Mioduszewski}}]{OrtizLeon2018}
{Ortiz-Le{\'o}n}, G.~N., {Loinard}, L., {Dzib}, S.~A., {et~al.} 2018, \apjl, 869, L33

\bibitem[{{Ortiz-Le{\'o}n} {et~al.}(2021){Ortiz-Le{\'o}n}, {Plunkett}, {Loinard}, {Dzib}, {Rodr{\'\i}guez-Garza}, {Pillai}, {Gong}, \& {Brunthaler}}]{Ortiz-Leon2021}
{Ortiz-Le{\'o}n}, G.~N., {Plunkett}, A.~L., {Loinard}, L., {et~al.} 2021, \aj, 162, 68

\bibitem[{{Patel} {et~al.}(2007){Patel}, {Curiel}, {Zhang}, {Sridharan}, {Ho}, \& {Torrelles}}]{Patel2007}
{Patel}, N.~A., {Curiel}, S., {Zhang}, Q., {et~al.} 2007, \apjl, 658, L55

\bibitem[{{Peck} \& {Impellizzeri}(2018)}]{Peck2018}
{Peck}, A.~B. \& {Impellizzeri}, C.~M.~V. 2018, in IAU Symposium, Vol. 336, Astrophysical Masers: Unlocking the Mysteries of the Universe, ed. A.~{Tarchi}, M.~J. {Reid}, \& P.~{Castangia}, 405--410

\bibitem[{{Pickett} {et~al.}(1998){Pickett}, {Poynter}, {Cohen}, {Delitsky}, {Pearson}, \& {M{\"u}ller}}]{Pickett1998}
{Pickett}, H.~M., {Poynter}, R.~L., {Cohen}, E.~A., {et~al.} 1998, \jqsrt, 60, 883

\bibitem[{{Planck Collaboration} {et~al.}(2020){Planck Collaboration}, {Aghanim}, {Akrami}, {Ashdown}, {Aumont}, {Baccigalupi}, {Ballardini}, {Banday}, {Barreiro}, {Bartolo}, {Basak}, {Battye}, {Benabed}, {Bernard}, {Bersanelli}, {Bielewicz}, {Bock}, {Bond}, {Borrill}, {Bouchet}, {Boulanger}, {Bucher}, {Burigana}, {Butler}, {Calabrese}, {Cardoso}, {Carron}, {Challinor}, {Chiang}, {Chluba}, {Colombo}, {Combet}, {Contreras}, {Crill}, {Cuttaia}, {de Bernardis}, {de Zotti}, {Delabrouille}, {Delouis}, {Di Valentino}, {Diego}, {Dor{\'e}}, {Douspis}, {Ducout}, {Dupac}, {Dusini}, {Efstathiou}, {Elsner}, {En{\ss}lin}, {Eriksen}, {Fantaye}, {Farhang}, {Fergusson}, {Fernandez-Cobos}, {Finelli}, {Forastieri}, {Frailis}, {Fraisse}, {Franceschi}, {Frolov}, {Galeotta}, {Galli}, {Ganga}, {G{\'e}nova-Santos}, {Gerbino}, {Ghosh}, {Gonz{\'a}lez-Nuevo}, {G{\'o}rski}, {Gratton}, {Gruppuso}, {Gudmundsson}, {Hamann}, {Handley}, {Hansen}, {Herranz}, {Hildebrandt}, {Hivon}, {Huang}, {Jaffe}, {Jones}, {Karakci}, {Keih{\"a}nen},
  {Keskitalo}, {Kiiveri}, {Kim}, {Kisner}, {Knox}, {Krachmalnicoff}, {Kunz}, {Kurki-Suonio}, {Lagache}, {Lamarre}, {Lasenby}, {Lattanzi}, {Lawrence}, {Le Jeune}, {Lemos}, {Lesgourgues}, {Levrier}, {Lewis}, {Liguori}, {Lilje}, {Lilley}, {Lindholm}, {L{\'o}pez-Caniego}, {Lubin}, {Ma}, {Mac{\'\i}as-P{\'e}rez}, {Maggio}, {Maino}, {Mandolesi}, {Mangilli}, {Marcos-Caballero}, {Maris}, {Martin}, {Martinelli}, {Mart{\'\i}nez-Gonz{\'a}lez}, {Matarrese}, {Mauri}, {McEwen}, {Meinhold}, {Melchiorri}, {Mennella}, {Migliaccio}, {Millea}, {Mitra}, {Miville-Desch{\^e}nes}, {Molinari}, {Montier}, {Morgante}, {Moss}, {Natoli}, {N{\o}rgaard-Nielsen}, {Pagano}, {Paoletti}, {Partridge}, {Patanchon}, {Peiris}, {Perrotta}, {Pettorino}, {Piacentini}, {Polastri}, {Polenta}, {Puget}, {Rachen}, {Reinecke}, {Remazeilles}, {Renzi}, {Rocha}, {Rosset}, {Roudier}, {Rubi{\~n}o-Mart{\'\i}n}, {Ruiz-Granados}, {Salvati}, {Sandri}, {Savelainen}, {Scott}, {Shellard}, {Sirignano}, {Sirri}, {Spencer}, {Sunyaev}, {Suur-Uski}, {Tauber}, {Tavagnacco},
  {Tenti}, {Toffolatti}, {Tomasi}, {Trombetti}, {Valenziano}, {Valiviita}, {Van Tent}, {Vibert}, {Vielva}, {Villa}, {Vittorio}, {Wandelt}, {Wehus}, {White}, {White}, {Zacchei}, \& {Zonca}}]{Planck2020}
{Planck Collaboration}, {Aghanim}, N., {Akrami}, Y., {et~al.} 2020, \aap, 641, A6

\bibitem[{{Plunkett} {et~al.}(2015){Plunkett}, {Arce}, {Mardones}, {van Dokkum}, {Dunham}, {Fern{\'a}ndez-L{\'o}pez}, {Gallardo}, \& {Corder}}]{Plunkett2015}
{Plunkett}, A.~L., {Arce}, H.~G., {Mardones}, D., {et~al.} 2015, \nat, 527, 70

\bibitem[{{Reid}(2007)}]{Reid2007}
{Reid}, M.~J. 2007, in Astrophysical Masers and their Environments, ed. J.~M. {Chapman} \& W.~A. {Baan}, Vol. 242, 522--529

\bibitem[{{Ronconi} {et~al.}(2024){Ronconi}, {Lapi}, {Torsello}, {Bressan}, {Donevski}, {Pantoni}, {Behiri}, {Boco}, {Cimatti}, {D'Amato}, {Danese}, {Giulietti}, {Perrotta}, {Silva}, {Talia}, \& {Massardi}}]{Ronconi2024}
{Ronconi}, T., {Lapi}, A., {Torsello}, M., {et~al.} 2024, \aap, 685, A161

\bibitem[{{Sadavoy} {et~al.}(2019){Sadavoy}, {Stephens}, {Myers}, {Looney}, {Tobin}, {Kwon}, {Commer{\c{c}}on}, {Segura-Cox}, {Henning}, \& {Hennebelle}}]{Sadavoy2019}
{Sadavoy}, S.~I., {Stephens}, I.~W., {Myers}, P.~C., {et~al.} 2019, \apjs, 245, 2

\bibitem[{{Santos} {et~al.}(2024){Santos}, {Enrique-Romero}, {Lamberts}, {Linnartz}, \& {Chuang}}]{Santos2024}
{Santos}, J.~C., {Enrique-Romero}, J., {Lamberts}, T., {Linnartz}, H., \& {Chuang}, K.-J. 2024, ACS Earth and Space Chemistry, 8, 1646

\bibitem[{{Slavicinska} {et~al.}(2025){Slavicinska}, {Boogert}, {Tychoniec}, {van Dishoeck}, {van Gelder}, {Navarro}, {Santos}, {Klaassen}, {Kavanagh}, \& {Chuang}}]{Slavicinska2025}
{Slavicinska}, K., {Boogert}, A.~C.~A., {Tychoniec}, {\L}., {et~al.} 2025, \aap, 693, A146

\bibitem[{{Sunada} {et~al.}(2007){Sunada}, {Nakazato}, {Ikeda}, {Hongo}, {Kitamura}, \& {Yang}}]{Sunada2007}
{Sunada}, K., {Nakazato}, T., {Ikeda}, N., {et~al.} 2007, \pasj, 59, 1185

\bibitem[{{Suutarinen} {et~al.}(2014){Suutarinen}, {Kristensen}, {Mottram}, {Fraser}, \& {van Dishoeck}}]{Suutarinen2014}
{Suutarinen}, A.~N., {Kristensen}, L.~E., {Mottram}, J.~C., {Fraser}, H.~J., \& {van Dishoeck}, E.~F. 2014, \mnras, 440, 1844

\bibitem[{{Urquhart}(2024)}]{Urquhart2024}
{Urquhart}, J.~S. 2024, in IAU Symposium, Vol. 380, Cosmic Masers: Proper Motion Toward the Next-Generation Large Projects, ed. T.~{Hirota}, H.~{Imai}, K.~{Menten}, \& Y.~{Pihlstr{\"o}m}, 135--151

\bibitem[{{van Kempen} {et~al.}(2009){van Kempen}, {Wilner}, \& {Gurwell}}]{vanKempen2009}
{van Kempen}, T.~A., {Wilner}, D., \& {Gurwell}, M. 2009, \apjl, 706, L22

\bibitem[{{van Terwisga} \& {Hacar}(2023)}]{Terwisga2023}
{van Terwisga}, S.~E. \& {Hacar}, A. 2023, \aap, 673, L2

\bibitem[{{Vastel} {et~al.}(2015){Vastel}, {Bottinelli}, {Caux}, {Glorian}, \& {Boiziot}}]{Vastel2015}
{Vastel}, C., {Bottinelli}, S., {Caux}, E., {Glorian}, J.~M., \& {Boiziot}, M. 2015, in SF2A-2015: Proceedings of the Annual meeting of the French Society of Astronomy and Astrophysics, 313--316

\bibitem[{{Volvach} {et~al.}(2022){Volvach}, {Volvach}, \& {Larionov}}]{Volvach2022}
{Volvach}, A.~E., {Volvach}, L.~N., \& {Larionov}, M.~G. 2022, \aj, 164, 66

\bibitem[{{Volvach} {et~al.}(2023){Volvach}, {Volvach}, \& {Larionov}}]{Volvach2023}
{Volvach}, A.~E., {Volvach}, L.~N., \& {Larionov}, M.~G. 2023, \aap, 672, A182

\bibitem[{{Voronkov} {et~al.}(2014){Voronkov}, {Caswell}, {Ellingsen}, {Green}, \& {Breen}}]{Voronkov2014}
{Voronkov}, M.~A., {Caswell}, J.~L., {Ellingsen}, S.~P., {Green}, J.~A., \& {Breen}, S.~L. 2014, \mnras, 439, 2584

\bibitem[{{Wakelam} {et~al.}(2004){Wakelam}, {Caselli}, {Ceccarelli}, {Herbst}, \& {Castets}}]{Wakelam2004}
{Wakelam}, V., {Caselli}, P., {Ceccarelli}, C., {Herbst}, E., \& {Castets}, A. 2004, \aap, 422, 159

\bibitem[{{Wang} {et~al.}(2023){Wang}, {Zhang}, {Yu}, {Wang}, {Yan}, {Chen}, {Zhao}, \& {Zou}}]{Wang2023}
{Wang}, Y.~X., {Zhang}, J.~S., {Yu}, H.~Z., {et~al.} 2023, \apjs, 264, 48

\bibitem[{{Wilking} \& {Claussen}(1987)}]{Wilking1987}
{Wilking}, B.~A. \& {Claussen}, M.~J. 1987, \apjl, 320, L133

\bibitem[{{Wilking} {et~al.}(1994){Wilking}, {Claussen}, {Benson}, {Myers}, {Terebey}, \& {Wootten}}]{Wilking1994}
{Wilking}, B.~A., {Claussen}, M.~J., {Benson}, P.~J., {et~al.} 1994, \apjl, 431, L119

\bibitem[{{Yang} {et~al.}(2023){Yang}, {Gong}, {Menten}, {Urquhart}, {Henkel}, {Wyrowski}, {Csengeri}, {Ellingsen}, {Bemis}, \& {Jang}}]{Yang2023}
{Yang}, W., {Gong}, Y., {Menten}, K.~M., {et~al.} 2023, \aap, 675, A112

\bibitem[{{Zeng} {et~al.}(2020){Zeng}, {Zhang}, {Jim{\'e}nez-Serra}, {Tercero}, {Lu}, {Mart{\'\i}n-Pintado}, {de Vicente}, {Rivilla}, \& {Li}}]{Zeng2020}
{Zeng}, S., {Zhang}, Q., {Jim{\'e}nez-Serra}, I., {et~al.} 2020, \mnras, 497, 4896

\end{thebibliography}

\appendix

\section{Individual sources}
\label{sec:sources}

The present sample consists of Galactic low-mass Class~0-Class~I YSOs for which water maser emission at 22~GHz was reported. As a summarized description for several of the sources studied here was presented previously by \citet{Humire2024}, we will briefly describe water-maser related information only. The exception is YLW16A, for which we provide more detailed information, as it was not covered in depth in our previous work due to the absence of CH$_{3}$OH maser emission.

NGC 2024 FIR~5/6 is located in the Orion B region, this star-forming region presents seven compact dust condensations, FIR~1--7, as resolved by \citet{Mezger1992}. Among these condensations, FIR~5 and 6 have been suggested to be Class~0 sources were water maser emission at 22~GHz emerges, although their exact emitting source among these two is not fully defined \citep[and references therein]{Furuya2003}. However, later studies focusing on the SED of some sub-clumps in NGC~2024~FIR~6 find that one of these highly-resolved sources corresponds to a hyper-compact (HC) H~\textsc{ii} region \citep{Choi2012}. As FIR~6 and 5 are separated by about 10$\arcsec$, less than our APEX beam of 19\farcs2 at 325~GHz, it may well be that this HC H~\textsc{ii} region, called FIR~6c, is dominating the SED of NGC 2024 FIR~5 (taken at 20$\arcsec$) at IR wavelengths, leading it to be the oldest in the studied sample.

CARMA~7 is a Class~0 protostar, it is the strongest radio source of the Serpens South protostellar cluster. It presents a bipolar and collimated outflow which extends $\sim$0.16\,pc north-south (PA$\sim$4$^{\circ}$ east of north) in CO emission \citep{Plunkett2015}. The outflow component, as seen by $^{12}$CO$J=$2--1, exhibits multiple knots with velocities exceeding 10~km~s$^{-1}$ and reaching up to 22~km~s$^{-1}$. The presence of MM is therefore likely associated with material moving at 5-10~km~s$^{-1}$ away from the source (and projected to $\sim$0.7~km~s$^{-1}$ from the LSR velocity of the system, as shown in Table~1 of \citet{Humire2024}), observed in about eight knots, namely, before dissociation (see below) \citep{Plunkett2015}. Recently, \citet{Ortiz-Leon2021} reported water maser emission at 22\,GHz in this source.

L1641N(orth) MM1/3 is a dark cloud in the Orion A region $\sim$7.2~pc south of the Orion Nebula Cluster, hosts the oldest stellar population in L1641 \citep{Terwisga2023}. Cloud-cloud collisions and protocluster winds, inferred from $^{12}$CO observations, may explain the CO shells centered at L1641N \citep{Nakamura2012}. Water (22~GHz) maser emission have been encountered in this source \citep{Kang2013}. Its protostar classification lies between Class~0 and I \citep{Froebrich2005}.

Serpens~FIRS~1 is a Class\,0 protostar located in the main core of the Serpens Molecular Cloud at a distance of 436$\pm$9\,pc \citep{OrtizLeon2018}. Also known as Serpens SMM\,1, this protostar is associated with a bipolar radio jet \citep[e.g.,][]{Curiel1996}. It is the most embedded, massive, and luminous YSO in the Serpens dark cloud \citep{Enoch2007} with a bolometric luminosity of 91\,$L_{\odot}$ \citep{Bae2011}, after correcting for the above-mentioned distance. High angular resolution observations at 0\farcs6 revealed a second YSO indicating a binary configuration for this system \citep{Choi2012}. Comparison between single dish \citep{Furuya2003} and interferometric observations reveal variable H$_{2}$O maser emission in this source \citep{Moscadelli2006}.

\begin{table}
\caption{Sources observed in our water maser survey.}
\setlength{\tabcolsep}{5pt}  
\label{tab:H2O_detections_coords_vlsr}
\centering
\begin{tabular}{llll} 
\hline \hline
Source & RA (J2000) & DEC (J2000) & $V_{\rm{LSR}}$ \\
& (h m s) & ($^\circ \,' \,''$) & (km s$^{-1}$) \\
\hline
IRAS 16293-2422 & 16 32 22.6 & -24 28 31.8 & 4.0 \\
L1641N\,MM1/3 & 05 36 18.4 & -06 22 11 & 7.4 \\
CARMA–7 & 18 30 04.1 & -02 03 02.4 & 7.0 \\
Serpens~FIRS~1 & 18 29 49.8 & +01 15 21 & 8.0 \\
YLW16A & 16 27 28.0 & -24 39 33.5 & 3.6 \\
NGC\,2071 North & 05 47 42.3 & +00 38 40 & 9.0 \\
Orion\,A\,West & 05 32 41.7 & -05 35 47.6 & 7.0 \\
HH1-2\,VLA1 & 05 36 22.5 & -06 46 01 & 8.9 \\
HH\,212 & 05 43 51.1 & -01 03 01 & 1.7 \\
Haro4-255 & 05 39 22.3 & -07 26 45 & 4.8 \\
GSS 30-IRS1 & 16 26 21.4 & -24 23 04.0 & 3.6 \\
NGC 2024 FIR 5/6 & 05 41 44.5 & -01 55 43 & 11.0 \\
L483 & 18 17 35.0 & -04 39 48 & 6.0 \\
L483-FIR & 18 17 29.8 & -04 39 38.3 & 6.0 \\
HOPS 96 & 05 35 30 & -04 58 48.8 & 11.7 \\
IRAS 18264-0143 & 18 29 05.3 & -01 41 56.9 & 7.0 \\
LDN 723-mm & 19 17 53.9 & +19 12 25 & 10.5 \\
Serpens\,SMM\,4 & 18 29 57.0 & +01 13 15.1 & 8.0 \\
\hline
\end{tabular}
\tablefoot{Sources of our survey. Their position are shown in columns 2 and 3. The LSR velocity is given in column 4.}
\end{table}

IRAS~16293--2422 is a well-studied Class~0 protostellar system \citep[e.g.,][]{Froebrich2005} known for its strong water maser activity. Located in the Lynds 1689N region of the $\rho$ Ophiuchus cloud at a distance of 141~pc \citep{Dzib2018}, this hierarchical multiple system consists of two main components, IRAS~16293--2422A and IRAS~16293--2422B, surrounded by an extended envelope of $\sim$8000~AU \citep{Crimier2010, Jacobsen2018}. IRAS~16293--2422A is divided into two compact sub-sources, A1 and A2.

The distance to this source \citep[141~pc;][]{Dzib2018} was determined using astrometry of H$_2$O masers observed at 22~GHz with the Very Long Baseline Array (VLBA). These masers are associated with outflows powered by this system, which, when compact, can reach maximum shifts of 12--16~km~s$^{-1}$, as seen in CO $J=$1--0, with respect to the velocity of the system \citep{Mizuno1990}. Interestingly, IRAS~16293--2422 has been routinely monitored over several decades, revealing multiple H$_2$O super flares. This phenomenon has been linked to increased activity in the binary system A. Additionally, variability in H$_2$O masers on timescales of months has been reported by \citet{Volvach2022}. The preferentially origin of long-term H$_2$O masers is suggested to be the proto-planetary disk of the A1 compact source \citep{Volvach2023}.

\begin{table*}
\caption{Detected water maser components per source and transition.}
\setlength{\tabcolsep}{3pt} 
\label{tab.compdetections}
\renewcommand{\arraystretch}{1.2}
\begin{tabularx}{\textwidth}{l 
  >{\centering\arraybackslash}p{1.6cm} >{\centering\arraybackslash}p{1.6cm} 
  >{\centering\arraybackslash}p{1.6cm} >{\centering\arraybackslash}p{1.6cm} 
  >{\centering\arraybackslash}p{1.6cm} >{\centering\arraybackslash}p{1.6cm}
  | >{\centering\arraybackslash}p{1.6cm} >{\centering\arraybackslash}p{1.6cm} |
  >{\centering\arraybackslash}X}
\toprule
\hline
\multirow{2}{*}{Source} 
& \multicolumn{2}{c}{183~GHz} 
& \multicolumn{2}{c}{321~GHz} 
& \multicolumn{2}{c}{325~GHz}
& \multicolumn{2}{c|}{22~GHz}
& \multirow{2}{*}{Ref.} \\
& $V_{\rm peak}$ [km\,s$^{-1}$] & $F_{\nu}$ [Jy] 
& $V_{\rm peak}$ [km\,s$^{-1}$] & $F_{\nu}$ [Jy] 
& $V_{\rm peak}$ [km\,s$^{-1}$] & $F_{\nu}$ [Jy] 
& $V_{\rm peak}$ [km\,s$^{-1}$] & $F_{\nu}$ [Jy] & \\
\midrule
IRAS16293--2422 
& \multicolumn{2}{c}{$V_\mathrm{range}$: [--8.5, 7.4]} 
& \multicolumn{2}{c}{$V_\mathrm{range}$: [2.6, 7.5]} 
& \multicolumn{2}{c}{$V_\mathrm{range}$: [--4.1, 9.2]} 
& \multicolumn{2}{c|}{$V_\mathrm{range}$: [--8, 10]} &  \\
& --5.9 & 15.9 &       &        &       &         & --2.4 & 154.0 & R1 \\
& --2.3 & 218.6 &      &        & --1.8 & 16.3   & 0.3   & 20.5  & R2 \\
& 0.8   & 31.6  &      &        & 1.2   & 497.4  & 1.5   & 50.2  &   \\
& 4.0   & 55.1  & 5.1  & 217.7 & 4.7   & 15.0   & 1.9   & 95.2  &  \\
&        &        &      &        & 7.3   & 18.0   & 2.6   & 79.8  & \\
&        &        &      &        &        &        & 3.3   & 41.2  & \\
&        &        &      &        &        &        & 5.4   & 293.6 & \\
&        &        &      &        &        &        & 5.9   & 77.4  & \\
&        &        &      &        &        &        & 7.3   & 12.7  & \\
\midrule
Serpens FIRS~1 
& \multicolumn{2}{c}{$V_\mathrm{range}$: [0.7, 7.1]} 
& \multicolumn{2}{c}{$V_\mathrm{range}$: [11.0, 18.4]} 
& \multicolumn{2}{c}{} 
& \multicolumn{2}{c|}{$V_\mathrm{range}$: [6, 25]} & R1 \\
& 3.9   & 210.0 &       &        &       &         & 6.4   & 8.4   & R3 \\
&        &        & 12.0 & 2.2   &       &         & 9.1   & 29.0  &   \\
&        &        & 15.8 & 11.1  &       &         & 15.6  & 2.1   &   \\
&        &        &      &        &       &         & 17.0  & 8.3   &  \\
&        &        &      &        &       &         & 18.1  & 14.9  &  \\
&        &        &      &        &       &         & 21.5  & 7.7   &  \\
&        &        &      &        &       &         & 22.9  & 3.0   &  \\
&        &        &      &        &       &         & 23.6  & 37.8  &  \\
&        &        &      &        &       &         & 24.1  & 8.6   &  \\
\midrule
YLW16A 
& \multicolumn{2}{c}{$V_\mathrm{range}$: [--19.1, 11.2]} 
& \multicolumn{2}{c}{} 
& \multicolumn{2}{c}{} 
& \multicolumn{2}{c|}{$V_\mathrm{range}$: [--16, 17]} & R1 \\
& --16.0 & 28.6 &       &        &       &         & --15.2 & 366.0 & R2 \\
& --12.2 & 23.4 &       &        &       &         & --10.5 & 1.1   &  \\
& --5.6  & 6.5  &       &        &       &         & 15.4  & 6.6   &  \\
& 1.2    & 6.7  &       &        &       &         & 16.3  & 39.1  &  \\
& 7.9    & 5.7  &       &        &       &         &       &       &   \\
\midrule
L1641N-MM1/3 
& \multicolumn{2}{c}{$V_\mathrm{range}$: [--4.8, 5.8]} 
& \multicolumn{2}{c}{} 
& \multicolumn{2}{c}{} 
& \multicolumn{2}{c|}{$V_\mathrm{range}$: [--8, 26]} & R3 \\
& --0.7 & 4.0  &       &        &       &         & --7.3 & 1.8   & R4  \\
& 3.8   & 18.6 &       &        &       &         & 5.2   & 9.0   &  \\
&        &        &      &        &       &         & 6.5   & 196.9 &  \\
&        &        &      &        &       &         & 7.3   & 17.0  &  \\
&        &        &      &        &       &         & 7.9   & 59.1  &  \\
&        &        &      &        &       &         & 8.6   & 23.1  &  \\
&        &        &      &        &       &         & 11.5  & 8.6   &  \\
&        &        &      &        &       &         & 23.8  & 8.8   &   \\
&        &        &      &        &       &         & 25.6  & 2.7   &   \\
\midrule
CARMA~7 
& \multicolumn{2}{c}{$V_\mathrm{range}$: [1.0, 11.0]} 
& \multicolumn{2}{c}{} 
& \multicolumn{2}{c}{} 
& \multicolumn{2}{c|}{$V_\mathrm{range}$: [10, 13]} & R5 \\
& 4.4   & 22.6 &       &        &       &         & 10.5  & 0.1   &   \\
& 8.7   & 20.1 &       &        &       &         &       &       &         \\
\bottomrule
\end{tabularx}
\tablefoot{
Velocities and flux densities refer to individual maser components identified through multi-Gaussian fits to the spectral profiles. $V_{\rm{range}}$ indicates the total velocity interval over which maser emission was detected (shaded blue regions in Fig.~\ref{fig:profiles}). 
For each transition, the subcolumns report the peak velocity and flux density of each Gaussian component, listed in order of increasing velocity. Typical $1\sigma$ uncertainties are $\sim$0.1~km~s$^{-1}$ in velocity and $\sim$10\% in flux density (median values).
Where possible, components from different (sub)mm transitions are placed on the same row when their peak velocities are similar (within a few km~s$^{-1}$). Blank entries indicate either non-detection or lack of a velocity match across transitions.
The 22~GHz maser properties  are drawn from previous studies across different epochs and resolutions: SerpensFIRS1 \citep{Furuya2003, Bae2011}, L1641N,MM1/3 \citep{Bae2011, Kang2013}, YLW16A \citep{Furuya2003, Sunada2007}, and IRAS16293--2422 \citep{Furuya2003, Sunada2007}. When multiple nearby components are reported, we bin them in 0.5~km~s$^{-1}$ intervals and average the velocity; for repeated observations, we adopt the mean peak flux density. The last column indicates the following references (Ref.): \citet{Furuya2003}, R2: \citet{Sunada2007}, R3: \citet{Bae2011}, R4: \citet{Kang2013}, and R5: \citet{Ortiz-Leon2021}.
}
\end{table*}

YLW16A, also known as IRS~44, has been proposed to be a protobinary system \citep{Allen2002}, although without a clear determination in the sub-mm regime, were \citet{Artur_de_la_Villarmois2019} found accreted shocked gas (up to $\sim$10~km~s$^{-1}$ with respect to the source's velocity, as seen in SO$_{2}$ $18_{4,14} \rightarrow 18_{3,15}$ transition) and a disk-like structure, inferred from SO$_{2}$ and H$^{13}$CO$^{+}$ species, respectively. This latter molecule provides, through position-velocity diagrams, an stellar mass of 1.2$\pm$0.1~$M_{\odot}$. The same authors also provide a couple of reasons why CH$_{3}$OH is not detected, even when SO and SO$_{2}$ do. One explanation is that methanol requires a large number density in the pre-shock gas to be released, which is not currently reached by the system. The second possibility is that methanol is released but later destroyed by dissociation at velocities larger than 10~km~s$^{-1}$, as this molecule is expected to survive at moderate velocities, between 3 and 10~km~s$^{-1}$ \citep{Suutarinen2014}. That limitation could also explain the lack of MM in this source (see Fig.~\ref{fig:evol_masers}).

This source has been classified as a Class~I YSO by \citet{Sadavoy2019} based on outflow detection, evidence of a dusty envelope or core, and its bolometric temperature. Additionally, water maser emission at 22~GHz has been reported with high variability, sometimes absent at certain epochs, at velocities between -15 and -10~km~s$^{-1}$ relative to the source's velocity system \cite{Wilking1987,Furuya2003}.

\section{SED modeling and data acquisition}
\label{appen.SED_modeling}

We performed the SED fitting of our six masing sources (see Table~\ref{tab.proposed_evol_sequence}) using GalaPy \citep{Ronconi2024}, an open-source tool developed in Python/C$++$ that covers wavelengths from X-ray to radio frequencies. GalaPy assumes a Chabrier Initial Mass Function (IMF) \citep{Chabrier2003} and a $\Lambda$CDM cosmological model with standard parameters \citep{Planck2020}.

GalaPy allows for the selection of different Star Formation History (SFH) models. In this study, we employed the In-Situ model, which has proven effective in predicting emission in galaxies of various types and epochs \citep{Ronconi2024}. This model enables tracking the evolution of gas, dust, and metallicity consistently with star formation evolution, ensuring coherence among derived physical parameters.

Regarding dust, GalaPy implements a two-component model that avoids assuming a fixed attenuation curve, deriving it instead from structural parameters. These two components represent (1) the molecular cloud phase associated with young stellar populations and (2) a diffuse medium further attenuating stellar emission. Both contribute to IR emission through two separate modified gray bodies.

Among the various free parameters of this dust model, we fixed certain known values and ranges to ensure physically meaningful results and enhance computational efficiency. Specifically, we set the total number of molecular clouds to \( N_\text{MC} = 1 \), assuming an isolated molecular cloud (MC), and constrained their dusty and molecular radii to a typical range of 10--100~pc. A complete list of initial conditions for our SED models is given in Table~\ref{tab.gmcs_GalaPy_constant_parameters} while the resulting diffuse dust ($T_{\rm{DD}}$) and molecular cloud ($T_{\rm{MC}}$) temperatures are provided in Table~\ref{tab.SED_results}. Other SED modeling results such as MC's ages, SFRs, and metallicities are not described as they are highly dependent on extraction apertures and photometry at optical and radio regimes, which is beyond the scope of this work.

\begin{table}[htbp]
\small
\setlength{\tabcolsep}{0.12cm}
\begin{tabular}{lll}
\hline \hline
Parameter & Value/Range & Brief description \\
\hline
redshift & 0.0                      & \\
age      & ([4,7], log)             & age of the MC \\
sfh.tau\_star &  ([4,7], log)       & characteristic timescale (yrs.)\\
ism.tau\_esc  & ([4,7], log)        & Stars' escape time from MC (yrs.)\\
sfh.psi\_max  & ([-2.0, 3], log)    & Maximum SFR ($M_{\rm{\odot}}$~yr$^{-1}$) \\
ism.R\_MC     &  ([0.0, 2], log)    & MC radius (pc) \\  
sfh.tau\_quench & 1e+20             & Star formation quenching (yrs.)\\
ism.f\_MC & ([0.0, 1.0], lin)         & MCs' fraction into the ISM\\
ism.norm\_MC & 100.0                & MCs' normalization factor\\
ism.N\_MC & 1.0                     & number of MCs\\
ism.dMClow & 1.3                    & Extinction index $<$100$\mu$\\
ism.dMCupp & 1.6                    & Extinction index $\gtrsim$100$\mu$\\
ism.norm\_DD & 1.0                  & Diffuse dust norm. factor\\
ism.Rdust & ([0.0, 2.0], log)         & Radius of the diffuse dust (DD, pc)\\
ism.f\_PAH & ([0.0, 1.0], lin)        & DD fraction radiated by PAH\\
ism.dDDlow & 0.7                    & DD extinction index $<$100$\mu$\\
ism.dDDupp & 2.0                    & DD extinction index $\gtrsim$100$\mu$\\
syn.alpha\_syn & 0.75               & Spectral index\\
syn.nu\_self\_syn & 0.2             & Self-absorption frequency (GHz)\\
f\_cal & ([-5.0, 0.0], log)           & Calibration uncertainty\\
\hline
\end{tabular}
\caption{GalaPy input parameter values or ranges common to the six masing sources studied in this work. The terms ``log'' and ``lin'' next to the ranges indicate logarithmic (log10) and linear values, respectively.}
\label{tab.gmcs_GalaPy_constant_parameters}
\end{table}

For a comprehensive list of all GalaPy parameters beyond those considered in this work, which is restricted to the In-Situ model, we refer the reader to Table~B.3 in \cite{Ronconi2024} and its online documentation\footnote{\url{https://galapy.readthedocs.io/en/latest/general/free_parameters.html}}.

\begin{table}
\caption{GalaPy's SED-fitting results for each source.}
\begin{tabular}{llll}
\hline \hline
Source & $T_{\rm{MC}}$ & $T_{\rm{DD}}$  & Cont. peak \\ 
                & [K]           & [K]            & [${\AA}$]  \\ \hline
IRAS~16293      & 39.60$_{-1.61}^{+1.78}$   & 16.92$_{-4.70}^{+10.20}$ & 1.30 \\
CARMA~7        & 23.82$_{-2.454}^{+3.785}$  & 29.56$_{-2.50}^{+2.89}$   & 1.74 \\
YLW16A        & 24.62$_{-8.72}^{+12.54}$    & 37.93$_{-11.14}^{+28.00}$ & 1.58\\
Serpens~FIRS  & 34.89$_{-7.01}^{+7.72}$     & 42.07$_{-17.61}^{+15.42}$ & 1.47\\
L1641N        & 51.87$_{-15.69}^{+4.95}$    & 48.493$_{-11.54}^{+6.18}$  & 1.00\\
NGC~2024      & 65.85$_{-23.18}^{+24.38}$   & 94.04$_{-44.80}^{+62.58}$ & 1.04\\ 
\hline
\end{tabular}
\tablefoot{$T_{\rm{MC}}$ is the molecular cloud temperature, $T_{\rm{DD}}$ is the diffuse dust temperature. Both temperatures come from different gray body fits (see blue and yellow fitting lines inside the SEDs in Fig.~\ref{Fig.SEDs}, respectively). The continuum peak (Cont. peak) is the wavelength at which the SED model peaks. Other parameters, such as stellar mass or bolometric luminosity are not taking into account as they can be strongly affected by the chosen apertures or source's variability and are also not required for our analysis.}
\label{tab.SED_results}
\end{table}

\begin{figure*}[!ht]
\centering
\includegraphics[width=0.99\textwidth, trim={0 0 0 0}, clip]{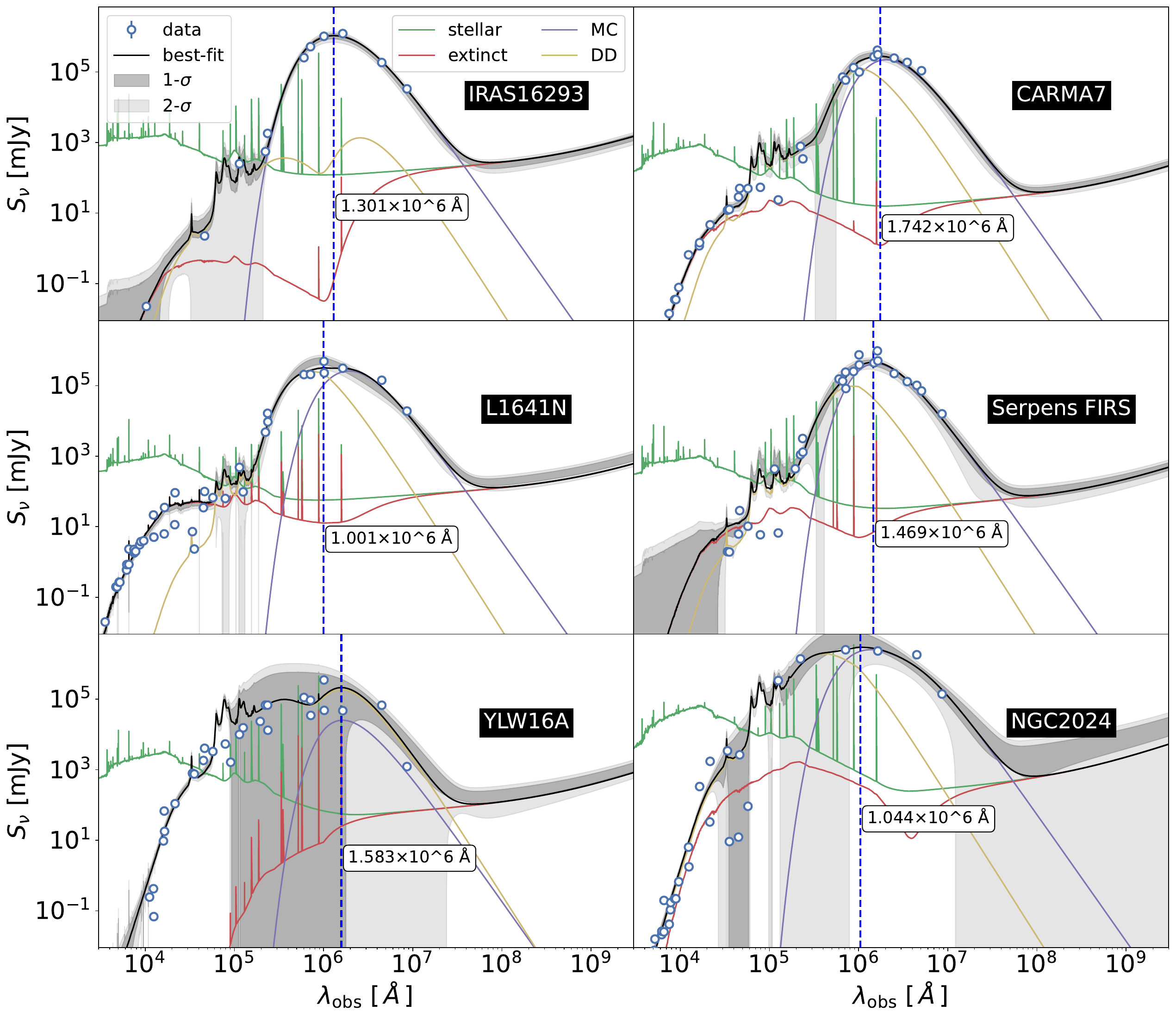}
\caption{SED models obtained with the GalaPy software for the sources studied in this work (see Appendix~\ref{sec:sources}). Points correspond to the photometric measurements obtained from the VizieR photometry viewer (see Appendix~\ref{appen.SED_modeling}), while solid lines correspond to the unattenuated stellar emission (green), molecular cloud component (MC, purple), stellar emission considering extinction (extinct, red), and diffuse dust (DD, yellow).}
\label{Fig.SEDs}
\end{figure*}

The photometric measurements for the six masing sources were obtained from the VizieR photometry viewer\footnote{\url{http://vizier.cds.unistra.fr/vizier/sed/old/}}, setting a conservative aperture radius of 20$\arcsec$ to better match our lower angular resolution of 34$\arcsec$, corresponding to the APEX half-power beam width at 181.310~GHz. Repeated observations were discarded leaving the highest in flux only. As we are interested in the dust temperature, quantity that depends on the FIR peak, we also deleted highly variable optical and radio measurements, while in the mid-IR we kept dispersed observations as the model naturally fit the largest values, as can be most noticeable seen for CARMA~7, L1641N~MM1/3, and Serpens~FIRS~1 in Fig.~\ref{Fig.SEDs}.

\section{Sulfur-bearing molecules as an age estimator}
\label{Appendix.sulfur}

As an age estimator, we avoid using sulfur-bearing molecules such as OCS, SO, and SO$_2$, which have been proposed as potential chemical clocks in previous studies (e.g., \citet{Wakelam2004, Herpin2009}). This decision is based on two key reasons:

\begin{enumerate}
    \item Limited observational constraints: Our measured line ratios (SO$J=$7$_{8}$--6$_{7}$/OCS$J=$16--15  $\sim$ 10--70 and SO$_2$$J=$9$_{1,9}$--8$_{0,8}$/OCS $\sim$ 1--8; not shown) show orders-of-magnitude differences compared to column density ratios from high-resolution ALMA observations of IRAS 16293-2422 \citep{Drozdovskaya2018}, where SO/OCS = 0.2\% (0.002) and SO$_2$/OCS = 0.5\% (0.005) in the compact component, and SO/OCS = 63\% (0.63) and SO$_2$/OCS = 9\% (0.09) in the more extended region. These discrepancies suggest our line ratios cannot be directly interpreted as column density ratios likely due to opacity effects mainly on the OCS line.

    \item Uncertainties in sulfur chemistry: The underlying chemical evolution of sulfur-bearing molecules remains poorly understood. Key issues include (i) the unknown reservoir of sulfur in interstellar ices (whether H$_2$S, atomic S, or other compounds like NH$_4$SH; \citealt{Slavicinska2025} are predominant), (ii) the sensitivity of sulfur chemistry to grain-surface processes and thermal history prior to protostar formation, and (iii) the strong dependence of observed abundances on desorption temperatures (e.g., SO$_2$ vs. OCS). Recent studies \citep{Santos2024} also highlight inconsistencies in using these species as evolutionary tracers, as their abundances may reflect local conditions (e.g., protostellar temperature) rather than source age.
\end{enumerate}

\end{document}